\documentstyle[twocolumn,epsbox,fodo98]{article}
\input{psfig}

\newtheorem{lemma}{Lemma}

\newtheorem{observation}{Observation}

\newcommand {\qed}{\vrule height4pt width4pt depth2pt}

\begin{document}
\bibliographystyle{alpha}

\raggedbottom
\def\BibTeX{{\rm B\kern-.05em{\sc i\kern-.025em b}\kern-.08em
    T\kern-.1667em\lower.7ex\hbox{E}\kern-.125emX}}

\title{\bf Efficient Retrieval of Similar Time Sequences Using DFT}
\author{{\bf Davood Rafiei}\\
        {\small drafiei@db.toronto.edu}\\
        {\small Department of Computer Science}\\
        {\small University of Toronto}
   \and {\bf Alberto Mendelzon}\\
        {\small mendel@db.toronto.edu}\\
        {\small Department of Computer Science}\\
        {\small University of Toronto}}
\date{}

\maketitle

\thispagestyle{empty}
\begin{figure}[b]
\begin{picture}(100,0)
\put(0,0){\line(1,0){100}}
\end{picture}
\\
\noindent
{\footnotesize\bf\raggedright
The 5th International Conference on Foundations of Data Organization
(FODO'98), Kobe, Japan, November 1998.
}
\end{figure}

\subsection*{\centering Abstract}
\noindent
{\it
We propose an improvement of the known
DFT-based indexing technique for fast retrieval of similar time sequences. 
We use the last few Fourier coefficients in the distance computation
without storing them in the index since
every coefficient at the end is the complex conjugate of a coefficient
at the beginning and as strong as its counterpart.
We show analytically that this observation can accelerate the search time of 
the index by more than a factor of two. 
This result was confirmed by our experiments,
which were carried out on real stock prices and synthetic data.
}

\paragraph{Keywords}
similarity retrieval, time series indexing

\section{Introduction}
\label{sec:intro}
Time sequences constitute a large amount of data stored in computers. Examples
include stock prices, exchange rates, weather data 
and bio\-me\-dic\-al measurements.
We are often interested in similarity queries on time-series data
\cite{shape95,Ag+95}.
For example, we may want to find stocks that behave in approximately the
same way; or years when
the temperature patterns in two regions of the world were similar.

There have been several efforts to develop access methods for
efficient retrieval of similar time sequences 
\cite{AgFaSw93,FRM94,RM97,YJF98}. 
Agrawal et al. \cite{AgFaSw93} propose an efficient index structure to
retrieve similar time sequences stored in a database. They map time sequences
into the frequency domain using the Discrete Fourier Transform (DFT)
and keep the first few coefficients in the index.
Two sequences are considered similar if their Euclidean distance is
less than a user-defined threshold.

In this paper, we propose using the last few Fourier coefficients
of a time sequence in the distance computation, the main observation being that,
every coefficient at the end is the complex conjugate of a coefficient
at the beginning and as strong as its counterpart.
This observation reduces the search time of the index by more than 50 percent
in most cases. 

The rest of the paper is organized as follows. In the next section we
review some background material on 
related work and on the discrete Fourier transform.
Our proposal on the efficient use of DFT in retrieving similar time 
sequences is discussed in Section~\ref{sec:store-retrieve}. 
In the same section, we present analytical results on the 
search time improvements of our proposed method.
Section~\ref{sec:expr} 
discusses the performance results obtained from experiments on real
and synthetic data. Section~\ref{sec:conclusion} is the conclusion.

\section{Background}
In this section, we briefly review background material on
past related work and on the discrete Fourier transform.

\subsection{Related Work}
There has been some follow-up work on the indexing technique proposed
by Agrawal et al. \cite{AgFaSw93}.
In an earlier work \cite{RM97}, we use this indexing method
and propose techniques for retrieving similar time sequences whose
differences can be removed by a linear transformation
such as moving average, time scaling and inverting.
In another work \cite{Raf-a98}, we generalize this framework to 
multiple transformations.
More follow-up work includes the work of Faloutsos et al. \cite{FRM94} 
on subsequence matching and that of 
Goldin et al. \cite{GoKa95} on normalizing sequences before storing them
in the index.
 
In this paper, we use the indexing technique proposed by
Agrawal et al.\cite{AgFaSw93}, but in addition to the first
few coefficients we also take the last few coefficients into account.
Both our analytical results and our experiments show that this observation 
accelerates the retrieval speed of the index by more than a factor of 2.
All follow-up works described earlier benefit from 
this performance improvement.

There are other related works on time series data.
A domain-independent framework for posing similarity queries on
a database is developed by Jagadish et al. \cite{JMM95}.
The framework has three components: a pattern language, a transformation 
rule language, and a query language. The framework can be tuned to 
the needs of time sequences domain.
Yi et al. \cite{YJF98} use time warping as a distance function
and present algorithms for retrieving similar time sequences under this
function. Agrawal et al. \cite{shape95} describe a pattern language called SDL
to encode queries about ``shapes'' found in time sequences.
A query language for time series data in the stock market domain is
developed by Roth \cite{mimsy}. The language is built on top of
CORAL \cite{RSS92}, and every query is translated into a sequence of
CORAL rules. Seshadri et al. \cite{SEQ94} develop a data model and
a query language for sequences in general but do not mention similarity
matching as a query language operator.

\subsection{Discrete Fourier Transform}
\label{sec:DFT}
Let a time sequence be a finite duration
signal $\vec{x}=[x_t]$ for $t=0,1, \cdots ,n-1$. The DFT
of $\vec{x}$, denoted by $\vec{X}$, is given by
\begin{equation}
X_f = \frac{1}{\sqrt{n}}\sum_{t=0}^{n-1} x_t e^{\frac{-j2\pi tf}{n}} \ \ \
f=0,1, \cdots, n-1
\label{eq:fft}
\end{equation}
where $j=\sqrt{-1}$ is the imaginary unit.
Throughout this paper, unless it is stated otherwise, we use small letters
for sequences in the
time domain and capital letters for sequences in the frequency domain.
The energy of signal $\vec{x}$ is given by
the expression
\begin{equation}
\label{eq:energy}
E(\vec{x}) = \sum_{t=0}^{n-1} |x_t|^2  .
\end{equation}

A fundamental observation that guarantees the correctness of the indexing
method for time series data is {Parseval's} rule \cite{Oppenheim75}, 
which states for 
a given signal $\vec{x}$ its energy remains the same after DFT, i.e.
\begin{equation}
\label{eq:parseval}
E(\vec{x}) = E(\vec{X}) 
\end{equation}
where $\vec{X}$ is the DFT of $\vec{x}$.
Using Parseval's rule and the linearity property of DFT 
(for example, see Oppenheim and Schafer \cite{Oppenheim75} for details), 
it is easy 
to show that the Euclidean distance between two signals in the time domain 
is the same as their distance in the frequency domain.
\begin{equation}
D^2(\vec{x},\vec{y}) = E(\vec{x} - \vec{y}) =
E(\vec{X} - \vec{Y}) = D^2(\vec{X},\vec{Y})
\label{eq:pars-eucl}
\end{equation}

\section{Storage and Retrieval of Similar Time Sequences}
\label{sec:store-retrieve}
Given a set of time series data, we can construct an index
(\cite{AgFaSw93}) as follows:
find the DFT of each sequence and keep the first few
DFT coefficients as the sequence features.
Let's assume that we keep the first {\it k} coefficients. Since all DFT
coefficients except the first one are complex numbers, keeping the first
$k$ DFT coefficients maps every time series into a point in a 
$(2k-1)$-dimensional space.
These points can be organized in a multidimensional index
such as R-tree family \cite{Gutt84,BKSS90} or grid files \cite{grid84}.
Keeping only the first {\it k} Fourier coefficients in the index
does not affect the correctness because the Euclidean distance between
any two points in the feature space is less than or
equal to their real distance due to Parseval's rule and the monotonic
property of the Euclidean distance.
Thus, the index always returns a superset of the answer set.
However, the performance of the index mainly depends on the energy
concentration of sequences within the first $k$ Fourier coefficients.
It turns out that a large class of real world sequences concentrate the energy
within the first few coefficients, i.e. they have a skewed energy spectrum
of the form $O(F^{-2b})$ for $b \geq 0.5$ where $F$ denotes the frequency.
For example, classical music and jazz fall in
the class of {\it pink noise} whose energy spectrum
is $O(F^{-1})$ (\cite{WSnoise90,Manfred91}), stock prices and exchange 
rates fall in the class of {\it brown noise} whose energy spectrum 
is $O(F^{-2})$ (\cite{Mandel77,Christopher84}), and the water level of 
rivers falls in the class of {\it black noise} for 
which $b>1$ (\cite{Mandel77,Manfred91}).

To retrieve similar time sequences stored in the index we 
may invoke one of the following spatial queries:
\begin{itemize}
\item {\bf Range Query:} Given a query point $\vec{Q}$ 
and a threshold $\epsilon$, find all points $\vec{X}$ such that
the Euclidean distance $D(\vec{X}, \vec{Q}) \leq \epsilon$.
\item {\bf Nearest Neighbor Query:} Given a query point $\vec{Q}$, 
find all points $\vec{X}$ such that
the Euclidean distance $D(\vec{X}, \vec{Q})$ is the minimum. Similarly,
a {\it k}-nearest neighbor query asks for the {\it k} closest points of a
given point.
\item {\bf All-Pair Query:} Given two multidimensional point sets
$s_1,s_2 \subseteq S$ and a threshold $\epsilon$,
find all pairs of points $(\vec{X},\vec{Y}) \in s_1 \times s_2$ such that
the Euclidean distance $D(\vec{X}, \vec{Y}) \leq \epsilon$.
\end{itemize}

\noindent Suppose we want to answer a range query using the index, i.e.,
to find all sequences $\vec{X}$ that are within
distance $\epsilon$ of a query sequence $\vec{Q}$, or equivalently
$D(\vec{X},\vec{Q}) < \epsilon$.
A common approach to answer this
query is to build a multidimensional rectangle of side $2\epsilon$
(or a multidimensional circle of radius $\epsilon$)
around $\vec{Q}$ and check for overlap between the query rectangle (circle)
and every rectangle in the index. That is, instead of checking
$D^2(\vec{X},\vec{Q}) < \epsilon^2$, we check
$|X_f - Q_f|^2 < \epsilon^2$
for $f=0, \ldots, k-1$. The latter is a necessary
(but not sufficient) condition for the former.

The size of the query rectangle has a strong effect on 
the number of directory nodes accessed  during the search process
and the number of candidates which includes
all qualifying data items plus some false positives
(data items whose full database records do not intersect the query region).
Our goal here is to reduce the size of the query region, 
using the inherent properties of DFT, without sacrificing the correctness.

\subsection{Our Proposal}
\label{sec:proposal}
The following lemma is central to our proposal.

\begin{lemma}
\label{lemma:DFT-sym}
The DFT coefficients of a real-valued sequence of duration $n$ satisfy
$X_{n - f} = X_f^*$ for $f=1, \ldots, n-1$ where the asterisk denotes
complex con\-ju\-ga\-tion\footnote{$(a+bj)^* = (a-bj)$}.
\end{lemma}

{\it Proof:} See Oppenheim and Schafer \cite[page 25]{Oppenheim75}. \qed

This means the Fourier transform of every real-valued
sequence is symmetric with respect to its middle.
A simple implication of this lemma is \\
$|X_{n - f}|\ =\ |X_f|$,
i.e. every amplitude at the beginning except the first one appears at the end.

\begin{observation}
In the class of (real-valued) time sequences that have an energy spectrum
of the form $O(F^{-2b})$ for $b \geq 0.5$, the DFT coefficients are not only
strong at the beginning but also strong at the end.
\end{observation}

This means if we do our distance computations based on only the 
first $k$ Fourier coefficients, we will miss all the information 
carried by the last $k$ Fourier coefficients which are as 
important as the former. However, the next observation shows that
the first $k$ Fourier coefficients are the only features that we need
to store in the index.

\begin{observation}
The first $\lceil(n+1)/2\rceil$ DFT coefficients of every (real-valued)
time sequence contain the whole information about the sequence.
\end{observation}

The point left to describe now is how we can take advantage of 
the last $k$ Fourier coefficients without storing them in the index.
We can write the Euclidean distance between two time sequences $\vec{x}$ 
and $\vec{q}$, using equations~\ref{eq:pars-eucl} and \ref{eq:energy}, 
as follows:
\begin{equation}
\label{pro:eq:euc}
D^2(\vec{x},\vec{q}) = D^2(\vec{X},\vec{Q}) =
\sum_{f=0}^{n-1} |X_f - Q_f|^2
\end{equation}
where $\vec{X}$ and $\vec{Q}$ are respectively DFTs of $\vec{x}$ and $\vec{q}$.
Since $|X_{n-f}|\ =\ |X_f|$
and $|Q_{n-f}|\ =\ |Q_f|$ for $f = 1,\ldots,n-1$,
we can write $D^2(\vec{X},\vec{Q})$ as follows:
\begin{eqnarray}
D^2(\vec{X},\vec{Q}) = |X_0 - Q_0|^2 + \nonumber \\
        \left\{
        \begin{array}{ll}
        \sum_{f=1}^{n/2 - 1} 2 |X_f - Q_f|^2 + \\
        \hspace{2em} |X_{n/2} - Q_{n/2}|^2 & for\ even\ n\\ \\
        \sum_{f=1}^{(n-1)/2} 2 |X_f - Q_f|^2 & for\ odd\ n\\
        \end{array}
        \right.	
\end{eqnarray}

A necessary condition for the left side to be less than $\epsilon^2$ is
that every magnitude on the right side be less than $\epsilon^2$.
For the time being and just for the purpose of presentation, 
we assume time sequences are normalized
\footnote{
A sequence is in normal form if its mean is 0 and its standard
deviation is 1.
} before being stored in the index. In general, time sequences
may be normalized because of efficiency reasons \cite{GoKa95} or 
other useful properties \cite{Raf-a98}.
Since the first Fourier coefficient is zero for normalized sequences,
there is no need to store it in the index.
In addition, since $k$ is usually a small number,
much smaller than $n$, we can assume that the $(n/2)th$ coefficient is also
not stored in the index. Now the condition left to be checked
on the index is
\[
2 |X_f - Q_f|^2 < \epsilon^2 \ \ \ for\ f=1, \ldots, k\\
\]
or, equivalently
\[
|X_f - Q_f| < \frac{\epsilon}{\sqrt{2}} \ \ \ for\ f=1, \ldots, k
\]

A common approach to check this condition is to build
a search rectangle of side $\frac{2\epsilon}{\sqrt{2}} = \sqrt{2} \epsilon$ 
(or a circle of diameter $\sqrt{2} \epsilon$) around $\vec{Q}$ and
check for an overlap between this rectangle (circle)
and every rectangle in the index. The search rectangle still guarantees 
to include all points within the Euclidean distance $\epsilon$ from
$\vec{Q}$, but there is a major drop in the number of false positives.
The effect of reducing the size of the search rectangle on the search time of
a range query is analytically discussed in the next section.

The symmetry property can be similarly used to reduce the size of 
the search rectangle even if sequences are not normalized.
The only difference is that one side of the search rectangle 
(the one representing the first DFT coefficient\footnote
{Note that the first DFT coefficient is a real number.}) 
is $2\epsilon$ and all other sides are 
$\sqrt{2} \epsilon$.

We can show that all-pair queries also benefit from the symmetry
property of DFT. Suppose we want to answer an all-pair query using two
R-tree indices, i.e., to find all pairs of sequences
that are within distance $\epsilon$ form each other.
A common approach for processing this query is to take pairs of
(minimum bounding) rectangles, one rectangle from each index, extend
the sides of one by $2\epsilon$ and check for a possible overlap with the other.
However, the symmetry property implies that if we extend every
side by $\sqrt{2} \epsilon$, the result is still guaranteed to include
all qualifying pairs though the number of false positives is reduced.

\subsection{Analytical Results on the Search Time Improvements}
There are two factors that affect the search time of a range query,
if we assume the CPU time to be negligible; one is the number of 
index nodes touched by the query rectangle and the other is
the number of data points inside the search rectangle (or candidates).
Both factors can be approximated by the area of the search rectangle,
if we assume data points are uniformly distributed over a unit square, 
and the search rectangle is a rectangle within this square
\footnote{We relax our assumptions later in this section.}.
Thus, to compare the search time of a rectangle of side 
$\sqrt{2} \epsilon$ to that of a one of side $2\epsilon$, 
we compare their areas.

Since a search rectangle has $2k$ sides, the area (or the volume)
of a search rectangle of side $\sqrt{2} \epsilon$ is 
$(\sqrt{2} \epsilon)^{2k} = 2^k\epsilon^{2k}$.
This is one $2^k$th of the area (or the volume) of a
rectangle of side $2\epsilon$ which is $(2\epsilon)^{2k} = 2^{2k}\epsilon^{2k}$.
Thus under the assumptions we have made, 
using a search rectangle of side $\sqrt{2} \epsilon$
instead of a one of side $2\epsilon$ should reduce the 
search time by $(1-1/2^k)*100$ percent. For example, using a
rectangle of side $\sqrt{2} \epsilon$ on an index built on 
the first two non-zero DFT coefficients should reduce 
the search time by 75 percent.

However, for the class of time sequences that have an
{\it energy spectrum} of the form $O(F^{-2b})$, 
the {\it amplitude spectrum} follows $O(F^{-b})$.
In particular for $b>0$, the amplitude reduces as a factor of frequency
and points get denser in higher frequencies.
If we assume that
the first non-zero DFT coefficient (for every data or query sequence) 
is uniformly distributed within a unit square, the $i$th DFT 
coefficient (for $i=1,\ldots,k$)
must be distributed uniformly within a square of side $i^{-b}$.
Thus keeping the first $k$ Fourier coefficients maps
sequences into points which are uniformly distributed within rectangle
$R=(<0,1>,<0,1>,<0,2^{-b}>$, $<0,2^{-b}>, \ldots, <0,k^{-b}>,<0,k^{-b}>)$.
 
In addition,
a search rectangle built on an arbitrarily chosen query point $\vec{Q}$
(inside or on $R$) is not necessarily contained fully within $R$.
If $\vec{Q}$ happens to be a central point of $R$,
the overlap between the two rectangles reaches its maximum.
We refer to this query as `the worst case query' since it requires
the largest number of disk accesses. On the other hand, if $\vec{Q}$ happens to
be a corner point of $R$, the overlap between the two rectangles
reaches its minimum. We call this query `the best case query'.
Thus the area of the overlap between the search rectangle and $R$, 
and as a result the search time, is not only a factor of $\epsilon$
but also a factor of $\vec{Q}$.

\begin{figure}[htbp]
\centering
\centerline{\psfig{figure=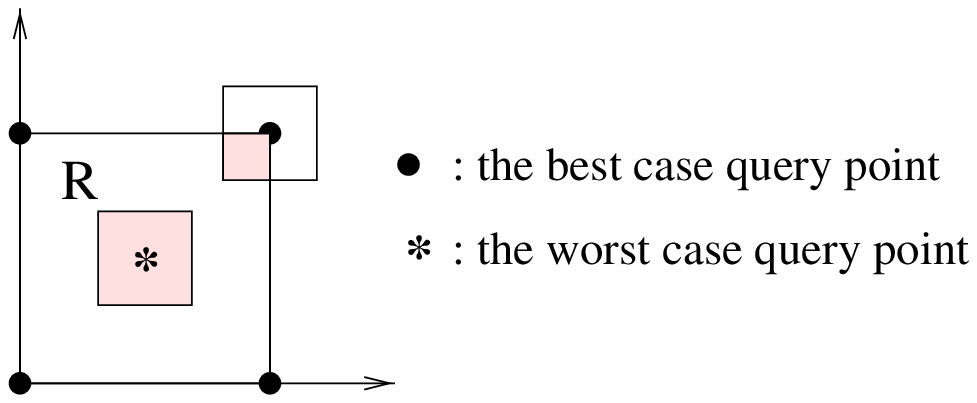,height=1.5in,width=3.2in}}
\label{fig:WBQ}
\end{figure}

To compare the search time of a query rectangle of side
$\sqrt{2} \epsilon$ to that
of one of side $2\epsilon$, we can compare their area of overlap with
$R$. 
The projection of the overlap between a search rectangle
of side $2\epsilon$ and $R$ to the $i$th DFT coefficient plane
is a square of side $min(i^{-b}, 2\epsilon)$ for the worst case query
and a square of side $min(i^{-b}, \epsilon)$ for the best case query.
Thus the area of the overlap between the search rectangle and $R$
for the worst case query is
$\prod_{i=1}^{k}(min(i^{-b}, 2\epsilon))^2$ 
and that for the best case query is
$\prod_{i=1}^{k}(min(i^{-b}, \epsilon))^2$.

To eliminate the effect of the size of $R$ in our estimates,
we divide the area of the overlap
by the area of $R$, i.e. $\prod_{i=1}^{k}(i^{-b})^2$, to get
what we call the {\it query selectivity}.
The query selectivity for the worst case query using a search rectangle of side
$2\epsilon$ can be expressed as follows:
\begin{eqnarray}
S(b,k,2\epsilon) 
	=\frac{\prod_{i=1}^{k}(min(i^{-b}, 2\epsilon))^2}
	   {\prod_{i=1}^{k}(i^{-b})^2}	\nonumber \\
	= \prod_{i=1}^{k}(min(i^{-b}, 2\epsilon)i^{b})^2.
\end{eqnarray}
The term $min(i^{-b}, 2\epsilon)i^{b}$
is 1 for $i^{-b} \leq 2\epsilon$ (or $i \geq (2\epsilon)^{-1/b}$) , and it is 
$2\epsilon i^{b}$ for $i^{-b} >2\epsilon$ (or $i < (2\epsilon)^{-1/b}$). 
Thus the query selectivity can be expressed as
\begin{equation}
S(b,k,2\epsilon) = \prod_{i=1}^{min(k,\lfloor(2\epsilon)^{-1/b}\rfloor)}
		   (2\epsilon i^{b})^2 
\end{equation}
It can be easily shown that $S(b,k,\epsilon)$ gives the query selectivity
for the best case query using the same search rectangle.
If we employ the symmetry property of the DFT, i.e. use
a search rectangle of side $\sqrt{2} \epsilon$, the query
selectivities for the  worst and the best case queries would be
$S(b,k,\sqrt{2} \epsilon)$ and $S(b,k,\frac{\epsilon}{\sqrt{2}})$ 
respectively.

\begin{figure}[htbp]
\centering
\centerline{\psfig{figure=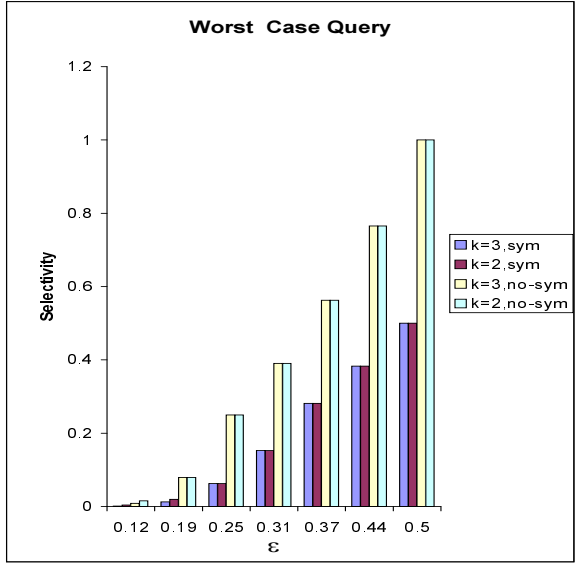,height=3.5in,width=3.5in}}
\caption{Query selectivity per search rectangle and {\it k} 
varying the threshold for the worst case query on Brownian noise data}
\label{fig:WQ}
\end{figure}

Figure~\ref{fig:WQ} shows the worst case query selectivity per 
search rectangle and $k$ varying the query threshold for Brownian noise 
data ($b=1$). As is shown, using the symmetry property reduces the query
selectivity by 50 to 75 percent for $k=2$ and $\epsilon \leq 0.5$.
If we keep the first three non-zero DFT coefficients ($k=3$), 
using the symmetry 
property reduces the selectivity by up to 87 percent.
In general, taking the symmetry property into account 
reduces the selectivity and as a result the search time in the 
worst case by 50 to $(1-1/2^k)*100$ percent for $k\geq 2$ 
and $\epsilon \leq 0.5$.

\begin{figure}[htbp]
\centering
\centerline{\psfig{figure=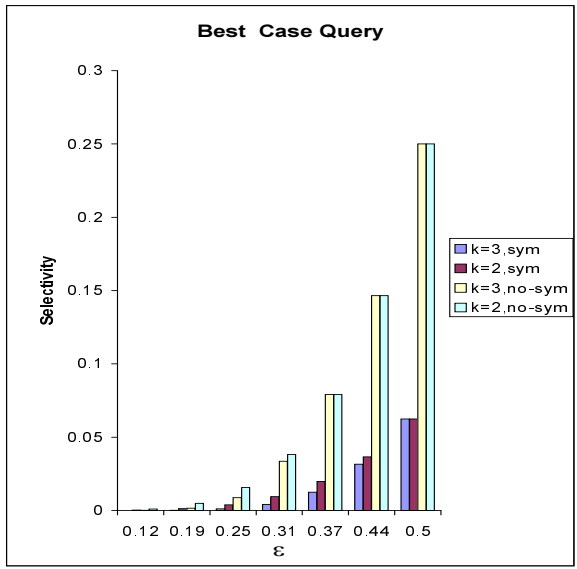,height=3.5in,width=3.5in}}
\caption{Query selectivity per search rectangle and {\it k}
varying the threshold for the best case query on Brownian noise data}
\label{fig:BQ}
\end{figure}

Figure~\ref{fig:BQ} shows the best case query selectivity per
search rectangle and $k$ varying the query threshold again for 
the Brownian noise data. 
As is shown, taking the symmetry property into account reduces
the selectivity by at least 75 percent for all values of $\epsilon \leq 0.5$, 
if we keep only the first two non-zero DFT coefficients. 
In general, taking the symmetry property into account 
reduces the selectivity and as a result the search time of the best 
case query by 75 to $(1-1/2^k)*100$ percent for $k\geq 2$ 
and $\epsilon \leq 0.5$.

\section{Experiments}
\label{sec:expr}
To show the performance gain of our proposed method, 
we implemented it using Norbert Beckmann's Version 2 implementation of the
R*-tree \cite{BKSS90} and compared it to 
the original indexing method proposed by Agrawal et al. \cite{AgFaSw93}.
All our experiments were conducted on a {168MHZ} Ultrasparc station.
We ran experiments on the following two data sets:
\begin{enumerate}
\item Real stock prices data obtained from the FTP site 
``ftp.ai.mit.edu/pub/stocks/results''. The data set consisted of 1067
stocks and their daily closing prices. Every stock had at least
128 days of price recordings.
\item Random walk synthetic sequences each of the form $\vec{x}=[x_t]$ where 
$x_t = x_{t-1} + z_t$ and $z_t$ is a uniformly distributed random number
in the range $[-500,500]$. The data set consisted of 20,000 sequences.
\end{enumerate}

We first transformed every sequence to its normal form, and 
then found its DFT coefficients. We kept the first $k$ DFT
coefficients as the sequence features. Since a DFT coefficient was a complex
number, a sequence became a point in a $2k$-dimensional space.
But the first DFT coefficient was always zero for normalized sequences,   
and we did not need to store it in the index; instead, we stored the mean 
and the standard deviation of a sequence along with its $k-1$ DFT
coefficients. In our experiments we used the polar representation
for complex numbers.

To do the performance comparison, we used both range and all-pair queries.
For range queries, we ran each experiment 100 times and each
time we chose a random query sequence from the data set and searched
for all other sequences within distance $\epsilon$ of the query sequence. 
We averaged the execution times from these runnings.
Our all-pair queries were spatial self-join queries where we searched the
data set for all sequence pairs within distance $\epsilon$ of each other.

\subsection{Varying the query threshold}
\begin{figure*}[htbp]
\psfig{figure=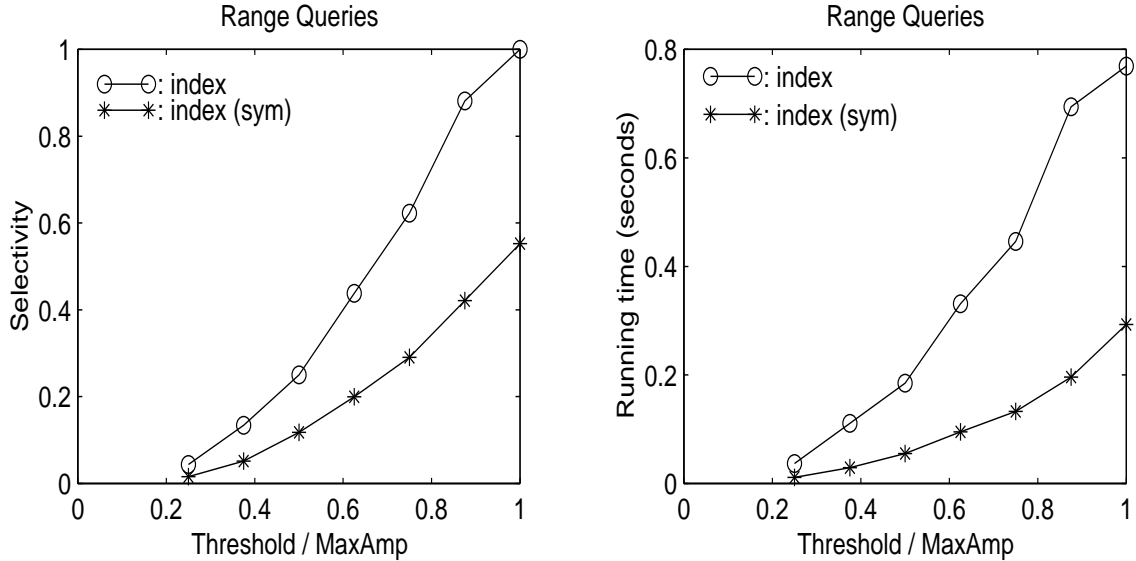,height=3in,width=6in}
\caption{Both query selectivities and running times for range queries 
varying the query threshold}
\label{fig:eps}
\end{figure*}

Our first experiment was on stock prices data consisting of 1067
time sequences each of length 128.
Our aim was to make a comparison between average case query selectivities 
obtained experimentally and the extreme case query selectivities
computed analytically.
We fixed the number of DFT coefficients to 2,
but we varied the query threshold from $1*MaxAmp$ to $0.24 * MaxAmp$
where $MaxAmp$ was the maximum amplitude of the first non-zero DFT 
coefficient over all sequences in the data set. Under this setting, 
a threshold $\epsilon * MaxAmp$ in our experiments was equivalent 
to threshold $\epsilon$ in our analytical results.
The average output size for $\epsilon=1*MaxAmp$ was 75 out of 1068 and that 
for $\epsilon=0.24* MaxAmp$ was zero, so we didn't try smaller thresholds.
Since query points were chosen randomly, we expected the query selectivity
for every threshold $\epsilon*MaxAmp$ to fall between the two extreme 
selectivities (the worst case and the best case) computed analytically 
for $\epsilon$. As is shown in Figure~\ref{fig:eps}, 
for $\epsilon / MaxAmp \leq 0.5$,
using the symmetry property reduces the query selectivity
by 53 to 64 percent and the search time by 70 to 74 percent.
It is consistent with our analytical results.
For $0.5 < \epsilon / MaxAmp \leq 1$, as the figure shows,
using the symmetry property reduces the query selectivity by 45 to 64
percent and the running time by 62 to 74 percent. 

\subsection{Varying the number of DFT coefficients}
\begin{figure*}[htbp]
\psfig{figure=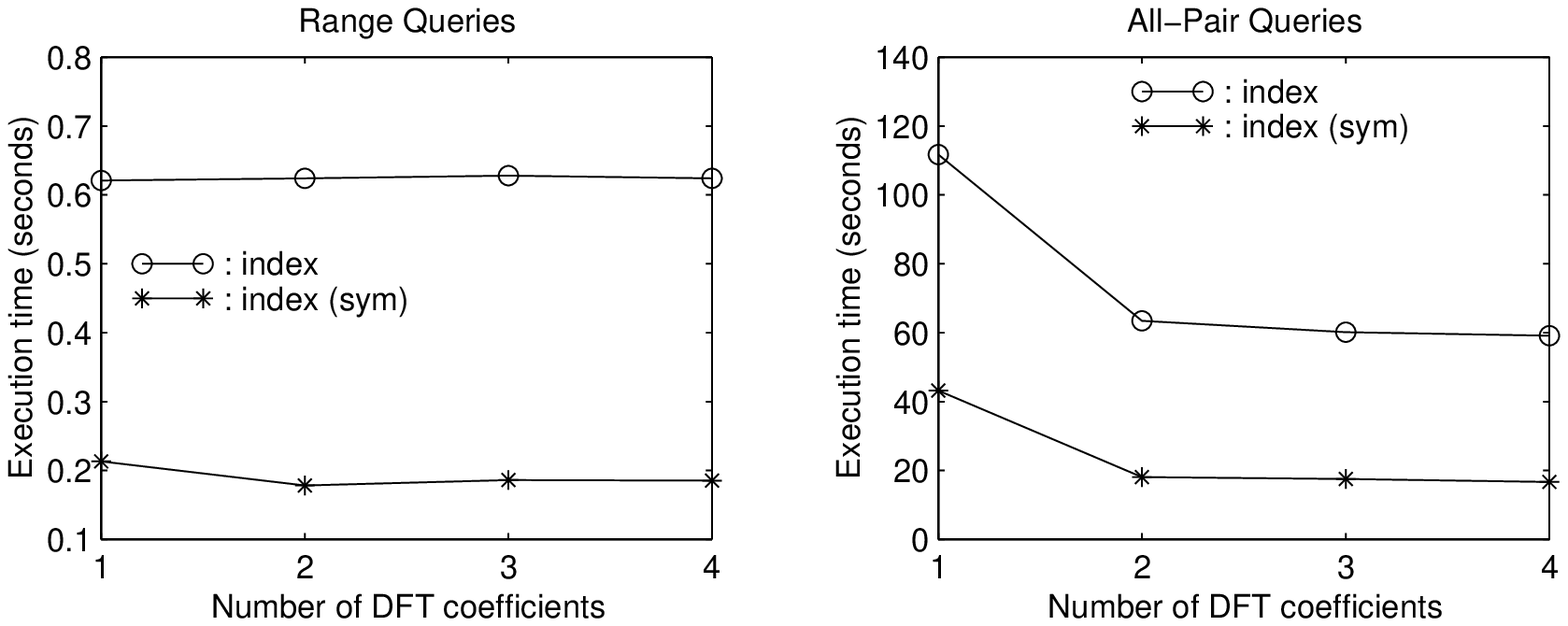,height=3in,width=6in}
\caption{Running times for range and all-pair queries varying the number 
of DFT coefficients}
\label{fig:varDFTq}
\end{figure*}

Our next experiment was again on stock prices data, but this time
we fixed the query threshold for range queries 
to $0.95 * MaxAmp$ and that for all-pair queries to $0.32 * MaxAmp$.
This setting gave us average output sizes of 30 and 203 respectively 
for range and all-pair queries.
We varied the number of DFT coefficients kept in the index from 1 to 4. 
Figure~\ref{fig:varDFTq} shows the running 
times per query for range and all-pair queries. 
Taking our observations into account reduces the search time of the
index by 66 to 72 percent for range queries and by
61 to 72 percent for all-pair queries.

\subsection{Varying the number of sequences}
\begin{figure*}[htbp]
\psfig{figure=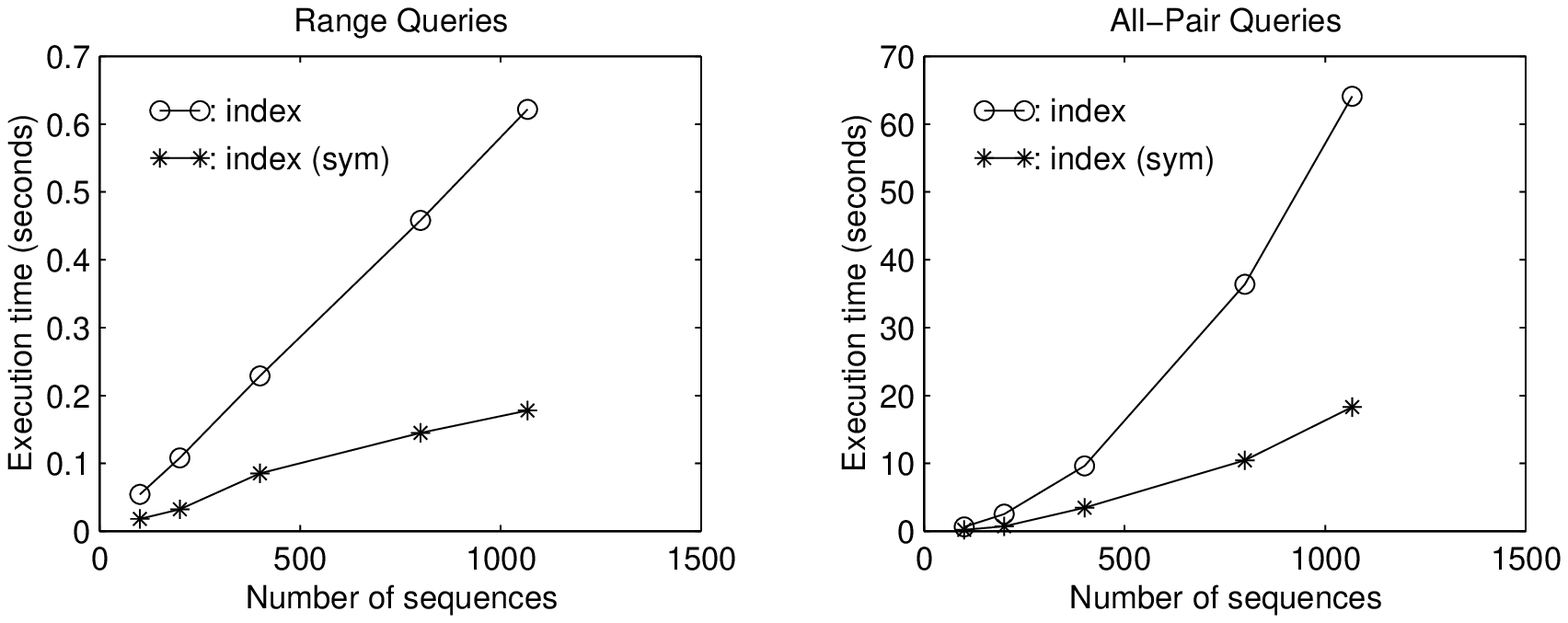,height=3in,width=6in}
\caption{Running times for range and all-pair queries varying 
the number of sequences}
\label{fig:varNoq}
\end{figure*}

In our next experiment, we fixed the number of DFT coefficient to
2 and the sequence length to 128, but we varied the number of sequences 
from 100 to 1067. The experiment conducted on stock prices data set. 
We again fixed the query threshold for range queries 
to $0.95 * MaxAmp$ and that for all-pair queries to $0.32 * MaxAmp$.
Figure~\ref{fig:varNoq} shows the running times 
per query for range and all-pair queries.
Our observation reduces the search time of the
index by 63 to 71 percent for range queries and by
64 to 72 percent for all-pair queries.

\subsection{Varying the length of sequences}
\begin{figure}[htbp]
\centering
\centerline{\psfig{figure=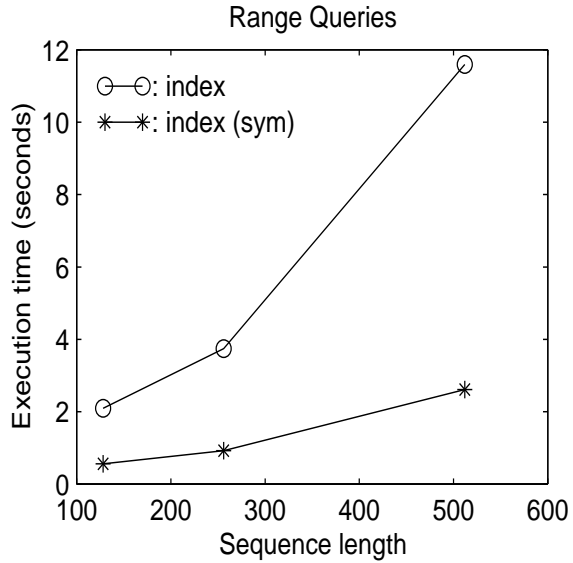,height=3in,width=3in}}
\caption{Running times for range queries varying the length of sequences}
\label{fig:varLenRq}
\end{figure}

Our last experiment was on synthetic data where
we fixed the number of DFT coefficients to
2 and the number of sequences to 20,000, but we varied the sequence
length from 128 to 512. The size of the data file was in the range
of 40 Mbytes (for sequences of length 128) to 160 Mbytes 
(for sequences of length 512). We fixed the query threshold to $0.44*MaxAmp$
and, based on our analytical results, we expected using
the symmetry property to reduce the search time by 50 to 75 percent.
Figure~\ref{fig:varLenRq} shows the running times 
per query for range queries.
Our proposed method reduces the search time of the
index by 73 to 77 percent. The search time improvement is slightly more than
our analytical estimates mainly because of the CPU time reduction
for distance computations which is not accounted for in our analytical 
estimates. Because of the high volume of data,
experiments on all-pair queries were very time consuming. 
For example, doing a self-join on sequences of length 512 did not 
finish after 12 hours of overnight running.
For this reason, we did not report them. 

\section{Conclusions}
\label{sec:conclusion}
We have proposed using the last few Fourier coefficients
of time sequences in the distance computation, the main observation being that
every coefficient at the end is the complex conjugate of a coefficient
at the beginning and as strong as its counterpart.
Our analytical observation shows that using the last few Fourier coefficients
in the distance computation accelerates the search time of
the index by more than a factor of two for a large range of thresholds. 
We also evaluated our proposed method over real and synthetic
data. Our experimental results were consistent with our analytical
observation; in all our experiments the proposed method reduced the
search time of the index by 61 to 77 percent for both range and
all-pair queries.

\section*{Acknowledgements}

\bibliography{ref}

\begin{thebibliography}{APWZ95}

\bibitem[AFS93]{AgFaSw93}
Rakesh Agrawal, Christos Faloutsos, and Arun Swami.
\newblock Efficient similarity search in sequence databases.
\newblock In {\em Foundations of Data Organizations and Algorithms (FODO)
  conference}, October 1993.

\bibitem[ALSS95]{Ag+95}
Rakesh Agrawal, King-Ip Lin, Harpreet~S. Sawhney, and Kyuseok Shim.
\newblock Fast similarity search in the presence of noise, scaling, and
  translation in time-series databases.
\newblock In {\em Proceedings of the 21st VLDB Conference}, pages 490--501,
  Zurich, Switzerland, 1995.

\bibitem[APWZ95]{shape95}
R.~Agrawal, G.~Psaila, E.~L. Wimmers, and M.~Zait.
\newblock Querying shapes of histories.
\newblock In {\em Proceedings of the 21st VLDB Conference}, pages 502--514,
  Zurich, Switzerland, 1995.

\bibitem[BKSS90]{BKSS90}
N.~Beckmann, H.-P. Kriegel, R.~Schneider, and B.~Seeger.
\newblock The {R* tree}: an efficient and robust index method for points and
  rectangles.
\newblock In {\em ACM SIGMOD Conf. on the Management Of Data}, pages 322--331.
  ACM, 1990.

\bibitem[Cha84]{Christopher84}
Christopher Chatfield.
\newblock {\em The Analysis of Time Series: an Introduction}.
\newblock Chapman and Hall, fourth edition, 1984.

\bibitem[FRM94]{FRM94}
C.~Faloutsos, M.~Ranganathan, and Y.~Manolopoulos.
\newblock Fast subsequence matching in time-series databases.
\newblock In {\em Intl. Conf. on Management of Data - SIGMOD 94}, pages
  419--429, Minneapolis, May 1994.

\bibitem[GK95]{GoKa95}
D.~Q. Goldin and P.~C. Kanellakis.
\newblock On similarity queries for time-series data: constraint specification
  and implementation.
\newblock In {\em 1st Intl. Conf. on the Principles and Practice of Constraint
  Programming}, pages 137--153. LNCS 976, Sept. 1995.

\bibitem[Gut84]{Gutt84}
Antonin Guttman.
\newblock R-trees: a dynamic index structure for spatial searching.
\newblock In {\em ACM SIGMOD Conf. on the Management Of Data}, pages 47--57.
  ACM, 1984.

\bibitem[JMM95]{JMM95}
H.~V. Jagadish, A.~O. Mendelzon, and T.~Milo.
\newblock Similarity-based queries.
\newblock {\em PODS}, 1995.

\bibitem[Man83]{Mandel77}
B.~Mandelbrot.
\newblock {\em Fractal Geometry of Nature}.
\newblock W.H. Freeman, New York, 1983.

\bibitem[NHS84]{grid84}
J.~Nievergelt, H.~Hinterberger, and K.~C. Sevcik.
\newblock The grid file: an adaptable, symmetric multikey file structure.
\newblock {\em ACM TODS}, 9(1):38--71, March 1984.

\bibitem[OS75]{Oppenheim75}
A.~V. Oppenheim and R.~W. Schafer.
\newblock {\em Digital Signal Processing}.
\newblock Prentice-Hall, Englewood Cliffs, N.J., 1975.

\bibitem[Raf98]{Raf-a98}
Davood Rafiei.
\newblock On similarity-based queries for time series data.
\newblock Submitted for publication, 1998.

\bibitem[RM97]{RM97}
Davood Rafiei and Alberto Mendelzon.
\newblock Similarity-based queries for time series data.
\newblock In {\em Proceedings of the ACM SIGMOD International Conference on
  Management of Data}, pages 13--24, Tucson, Arizona, May 1997.

\bibitem[Rot93]{mimsy}
William~G. Roth.
\newblock {MIMSY}: A system for analyzing time series data in the stock market
  domain.
\newblock University of Wisconsin, Madison, 1993.
\newblock Master Thesis.

\bibitem[RS92]{RSS92}
Raghu Ramakrishnan and Divesh Srivastava.
\newblock {CORAL}: Control, relations and logic.
\newblock In {\em Proceedings of the Int. Conf. on VLDB}, 1992.

\bibitem[Sch91]{Manfred91}
Manfred Schroeder.
\newblock {\em Fractals, Chaos, Power Laws: Minutes from an Infinite Paradise}.
\newblock W.H. Freeman, New York, 1991.

\bibitem[SLR94]{SEQ94}
P.~Seshadri, M.~Livny, and R.~Ramakrishnan.
\newblock Sequence query processing.
\newblock In {\em Proceedings of the ACM SIGMOD International Conference on
  Management of Data}, pages 430--441, 1994.

\bibitem[WS90]{WSnoise90}
B.J. West and M.~Shlesinger.
\newblock The noise in natural phenomena.
\newblock {\em American Scientist}, 78:40--45, Jan-Feb 1990.

\bibitem[YJF98]{YJF98}
Byoung-Kee Yi, H.~V. Jagadish, and Christos Faloutsos.
\newblock Efficient retrieval of similar time sequences under time warping.
\newblock In {\em Int. Conf. on Data Engineering}, 1998.

\end{thebibliography}
\end{document}